\documentclass[pra,twocolumn,nofootinbib,eqsecnum,floatfix,superscriptaddress]{revtex4}
\usepackage{epsfig}
\usepackage{bm}
\usepackage{amsmath}
\input epsf
\numberwithin{equation}{section}
\renewcommand{\theequation}{\arabic{section}.\arabic{equation}}

\def\ch{{\cal H}}
\def\cH{{\cal H}}

\def\cO{\hat{\cal O}}
\def\chO{{\hat{\cal O}}}
\def\dm{{\rm dim}}
\def\me{{\rm meas} \ }

\def\Ph{\hat P}
\def\Psih{\hat\Psi}
\def\Ch{\hat C}
\def\Hc{{\cal H}}
\def\H3s{^s{\cal H}^{3}}
\def\vy{{\vec y}}
\def\ep{\epsilon}
\def\pv{\pi^i}
\def\num{\nu}
\def\be{\begin{equation}}
\def\ee{\end{equation}}
\def\Pb{{\hat P}^{k\alpha_{k-1}\cdots\alpha_1}_{\alpha_k}}
\def\Rh{{\hat R}}
\def\R4{{\Rh^{4}_{\alpha_3\alpha_2\alpha_1}}}
\def\R3{{\Rh^{3}_{\alpha_2\alpha_1}}}

\def\branch{{\alpha_{k-1}\cdots\alpha_1}}
\def\branchk{{k\branch}}

\def\uf{}
\def\cf{}

\newcommand{\atwo}{\alpha_2 \alpha_1}
\newcommand{\atwop}{\alpha'_2\alpha'_1}
\newcommand{\athree}{\alpha_3\alpha_2\alpha_1}

\newcommand{\utwo}{2 \alpha_1}

\newcommand{\uthree}{3\alpha_2\alpha_1}
\newcommand{\uthreep}{3\alpha'_2\alpha'_1}

\def\be{\begin{equation}}
\def\ee{\end{equation}}

\begin{document}

\title{Adaptive Coarse Graining, Environment, Strong  Decoherence, \\
and Quasiclassical Realms}

\author{Murray Gell-Mann}

\email{mgm@santafe.edu}

\affiliation{Santa Fe Institute, Santa Fe, NM 87501}

\author{James B.~Hartle}

\email{hartle@physics.ucsb.edu}

\affiliation{Santa Fe Institute, Santa Fe, NM 87501}
\affiliation{Department of Physics, University of California,
 Santa Barbara, CA 93106-9530}

\date{\today }

\begin{abstract}
Three ideas are introduced that when brought together characterize the realistic quasiclassical realms of our quantum universe as particular kinds of sets of alternative coarse-grained histories defined by quasiclassical variables: (1) Branch dependent adaptive coarse grainings that can be close to maximally refined and can simplify calculation. (2) Narrative coarse grainings that describe how features of the universe change over time and allow the construction of an environment. (3)  A notion of strong decoherence that characterizes realistic mechanisms of decoherence. 
\end{abstract}



\maketitle

\section{Introduction}

A striking feature of our quantum universe is the wide range of time, place, and scale on which the deterministic laws of classical physics hold to an excellent approximation. What is the origin of this classical predictability in a quantum mechanical universe characterized fundamentally by indeterminacy and distributed probabilities? 

This paper is one of a series \cite{GH90a,GH90b,Har91a,GH93a,GH93b,Gel94,GH94,Har94b,Har09,GH95,GH07}
aimed at  characterizing the realistic quasiclassical realm(s) of our quantum universe as particular kinds of decoherent sets of coarse grained alternative histories defined by quasiclassical variables.
To this end we introduce three new (connected) ideas: branch dependent adaptive coarse grainings, a general notion of a narrative set of alternative histories, and a notion of strong decoherence to characterize realistic mechanisms of decoherence. We have discussed some of these before but this discussion we believe is simpler, more general, and more connected. The advantages and importance of these new ideas are as follows:

{\it Branch Dependent Adaptive Coarse Grainings:} 
   These allow for the possibility of coarse grainings that are close to maximal --- as refined as possible consistent with decoherence and  classicality.
 That way the quasiclassical realms can be a property of our universe, and not  just our choice. 
This kind of coarse graining can  simplify calculation and act against premature filling of the Hilbert space by not following low probability branches. 

{\it Narratives Sets of Histories:} These give a general characterization of a coarse graining whose histories tell a story about how features of the the universe change over time. They also allow the construction of an environment. They therefore put the notion of  environment in its proper place as a consequence of a narrative coarse graining, and not as a separate postulate of quantum mechanics. 

{\it Strong Decoherence:}   Strong decoherence is a  more realistic, but still general, notion of decoherence that characterizes realistic mechanisms of decoherence where records are created in variables other than those followed. {\uf The orthogonality of these records produces decoherence. }Strong decoherence ensures that the past remains permanent as a set of histories is fine-grained by extending it into the future.  

The paper is structured as follows: Section \ref{histories} reviews decoherent histories quantum mechanics, largely to establish our notation. Branch dependent coarse grainings and adaptive coarse grainings are described in Section \ref{branchcg}. The notion of narrative framework is introduced in Section \ref{narrative} and the system-environment split that follows from it in Section \ref{environments}. 
Section \ref{strong} gives a simplified account strong decoherence and Section \ref{features} is concerned with the notions of records and density matrices that follow from it.  
There are brief conclusions in Section \ref{conclusion}.

\section{Histories, Coarse Grainings and Decoherence}
\label{histories}\label{section2}

Largely to establish notation,  we  give a very brief account of some essential elements of the modern synthesis of ideas characterizing  the quantum mechanics of closed systems that we call decoherent histories quantum mechanics\footnote{For more detail see the classic expositions of Griffiths \cite{Gri02}, Omn\`es \cite{Omn94}, and Gell-Mann \cite{Gel94}.}.

\subsection{A Model Closed Quantum System}
We consider a closed quantum system, most generally the
universe, in the approximation that gross quantum fluctuations in the geometry of spacetime
can be neglected\footnote{For the generalizations that are needed for quantum spacetime see
{\it e.g.} \cite{Har95c,Har06}.}.  The closed system can then be thought of as a large (say
$\gtrsim$ 20,000 Mpc), perhaps expanding box of particles and fields in a 
fixed background spacetime.
Everything is contained within the box,  galaxies, planets, observers and
observed, measured subsystems, any apparatus that measures
them, and, in particular, any human observers including us.  This is a manageable model  of the most general physical context for prediction.

{\uf There is a Hilbert space $\cH$ for the contents of the box.}
The essential theoretical inputs to the process of prediction are
the Hamiltonian $H$ governing evolution and the  quantum state of the universe which we assume to be a pure\footnote{From the perspective of the time-neutral formulation of quantum theory (e.g. \cite{GH93b}), we are assuming also a final condition of ignorance.}  $|\Psi\rangle$.

\subsection{Histories}

We will work in the Schr\"odinger picture where one  operator represents the same quantity at all times. Operators in the Schr\"odinger picture will be distinguished from Heisenberg picture operators  by hats, viz.  $\hat{\cal O}$.

Sets of yes/no alternatives at one moment of time $t$ are 
represented by an exhaustive set of orthogonal projection operators 
$\{\Ph_\alpha\}$,
$\alpha = 1, 2, 3 \cdots$ satisfying
\begin{equation}
\sum\nolimits_\alpha \Ph_\alpha = I, \quad   \Ph_\alpha\, \Ph_\beta =
\delta_{\alpha\beta} \Ph_\alpha \ .
\label{twoone}
\end{equation}
These conditions  ensure that the projections represent an exhustive set of exclusive alternatives. 
A completely fine-grained description of a quantum system at a moment of time is provided by a set of one dimensional projections. All other sets are coarse-grained. 

The most general objective of quantum theory is the prediction of the probabilities of
individual members of sets of  alternative coarse-grained time histories of the closed
system. For instance, we might be interested in alternative histories of the center-of-mass
of the Earth in its progress around the Sun, or in histories of the correlation between the
registrations of  a measuring apparatus and a measured property of a subsystem. Histories are essential for defining classical behavior. For example, we say that the Earth moves in a classical orbit when the probability from $H$ and $\Psi$  is high for a history of  motion that is correlated in time by Newton's laws. 

An important kind of set of alternative histories is specified by sets of alternatives at a sequence of times $t_1<t_2<\cdots < t_n$. An individual history $\alpha$ in such a set is a particular sequence of alternatives  $\alpha \equiv (\alpha_1, \alpha_2, \cdots, \alpha_n)$ at the times\footnote{More generally alternatives can be extended over time \cite{STalt}. }  $t_1<t_2<\cdots < t_n$. Such a set of histories has a branching structure in which a history up to any given time $t_m \le t_n$ branches into further alternatives at later times. {\uf Each history is a `branch' of the branching structure.} 

Individual histories are represented by the  chains $\Ch_\alpha$ of the projection operators that define the alternatives  $\alpha=(\alpha_1, \alpha_2, \cdots, \alpha_n)$ at the times $t_1<t_2<\cdots < t_n$ with unitary evolution between times specified by  the Hamiltonian. The simplest examples are sets of histories defined by the same sets  alternatives $\{P_\alpha\}$  at  the series of times.  These histories are represented by the operators 
\begin{align}
\label{classop-simp}
\Ch_{\alpha} \equiv&\Ph_{\alpha_n}e^{-iH(t_{n}-t_{n-1})/\hbar}\Ph_{\alpha_{n-1}}e^{-iH(t_{n-1}-t_{n-2})/\hbar} \nonumber \\ &\cdots\Ph_{\alpha_1}e^{-iH(t_1-t_0)/\hbar} .
\end{align}
where $t_0<t_1$ is an initial time where the initial state of the box $\Psih(t_0)$ is specified. 

For any individual history $\alpha$ its {\it branch state vector}  at time $t_n$ is  defined by
\begin{equation}
\Psih_\alpha(t_n) =\Ch_\alpha \Psih(t_0) .
\label{twofive}
\end{equation}
The branch state vector $\Psih_\alpha(t)$ can be defined at any other time by  the unitary evolution of \eqref{twofive} with $H$. 
Evidently from \eqref{twoone} 
\be
\label{sumbranch}
\Psih(t) = \sum_\alpha \Psih_\alpha(t) . 
\ee

\subsection{Decoherence}

When probabilities can be consistently assigned to the individual histories in a set,
they are given by
\begin{equation}
p(\alpha) = \parallel \Psih_\alpha(t)\parallel^2 = 
\parallel \Ch_\alpha \Psih(t_0)\parallel^2\, .
\label{twosix}
\end{equation}
But because of quantum interference,  probabilities cannot be consistently assigned to every set of alternative
histories that may be described.  
Negligible interference between the branches of a set
\begin{equation}
(\Psih_\alpha(t), \Psih_\beta(t))\approx 0 \quad , \quad \alpha\not=\beta \, ,
\label{twoseven}
\end{equation}
is a sufficient condition for the probabilities \eqref{twosix} to be consistent with the
rules of probability theory. The orthogonality of the branches is approximate in
realistic situations. But we mean by \eqref{twoseven} equality to an accuracy that defines probabilities well beyond the standard to which they can be checked or the physical situation modeled. Sets of histories for which the interference is negligible according to \eqref{twoseven} are said to medium decohere.   Medium
decoherence is the weakest of known conditions that are consistent with elementary notions of the independence of isolated systems \cite{Dio04}.  

For characterizing quasiclassical realms stronger notions of decoherence characterizing realistic mechanisms of decoherence can be useful \cite{GH95}. A wide class of stronger notions replaces \eqref{twoseven} with 
\be
\label{strongerdec}
(\Psih_\alpha(t), {\cal O}\Psih_\beta(t))\approx 0 \quad , \quad \alpha\not=\beta \, ,
\end{equation}
for some class of operators $\cal O$ including the identity. The notion of strong decoherence that we introduced in \cite{GH95} is an example {\uf that we will discuss in Section \ref{strong}}.

\subsection{Quasiclassical Variables}
\label{qcvariables}

Quasiclassical realms are defined by coarse grainings based on quasiclassical variables.  These are averages over small volumes of densities of approximately conserved quantities such as energy, momentum and numbers, such as baryon number, in epochs when they are conserved. The approximate conservation of these quantities is the source of their classical predictability in the face of the noise that typical mechanisms of decoherence produce (see, e.g. \cite{GH90a,GH93a,GH07,Hal98,Hal99,Har09}). 

{\uf A quasiclassical coarse-graining is specified by three things: (1) First,  a sequence of time steps $t_1,t_2, \cdots t_n$. (2) Second a partition of space into volumes $V(\vy)$ labeled by a triple of integers $\vy$.   (3)  
And third an exhaustive set of exclusive ranges of coarse grained values $\{\Delta_\beta\}$ for the averages over each volume $\vy$ of each of the quantities energy density, momentum density, and number density, at each time step.
A particular history is represented by sequences of  particular ranges $\Delta_\beta$ at each time step, for each volume, for each variable.   The construction of the relevant projection operators is described in more detail in Appendix \ref{appA}.}

As described here quasiclassical coarse grainings are conceptually simple but notationally messy. We will therefore use a highly condensed notation for them. We denote by $\Ph^k_{\alpha_k}$ the projections (or products of nearby projections) at each time step.
The superscript $k$ denotes the time step {\it and} the coarse graining at that time step. That is, it stands for the time step and the coarse graining ingredients  (2) and (3)  mentioned above at that time step. The index $\alpha_k$ denotes the particular alternative at the time step $k$. That is, it stands for the particular ranges $\Delta_\beta$, for each quasiclassical variable, in each {\uf volume} $\vy$.

\section{Branch Dependent Coarse-Grainings}
\label{branchcg}
This section develops the idea of branch dependent coarse grainings in which the {\it set} of alternatives defining the branches at one time depends on the {\it specific history} (branch) that preceded it.  The idea of adaptive branch dependent coarse grainings is also introduced in which coarse grainings are adapted to changing physical situations. Sets of histories that  describe realistic physical situations in the universe are almost always branch dependent.  Adaptive coarse grainings are the  efficient way of exhibiting the interesting features of these situations. We begin with two illustrative examples --- {\uf the first on the scale of the laboratory, the second on the scale of the cosmos.}

\subsection{Examples of Branch Dependence}

\subsubsection{A model measurement situation}
\label{measurement}

Imagine
a closed system consisting of  an isolated laboratory containing an experimenter. The lab is  equipped with apparatus for measuring the spin of an isolated electron in a prepared state. The apparatus can be adjusted to measure the spin along any direction the experimenter chooses. The experimenter, the apparatus, and the electron are all quantum mechanical physical systems within the closed laboratory. At time $t_1$ the experimenter flips a coin (or uses a quantum random bit generator) to decide  whether to measure the spin along the $x$-axis or $z$-axis. Heads it's the $z$-direction; tails it's the $x$-direction.  At time $t_2$ she carries the measurement out.  The relevant histories for describing this situation consist of chains of projections at two times. The alternatives at the first time describe the direction chosen for the spin measurement, $(\Ph_{\me x}, \Ph_{\me z})$ in what we hope is an obvious notation.  At the second time the alternatives are $(\Ph^z_+, P^z_-)$ if first alternative was $\Ph_{\me z}$ and $(\Ph^x_+, \Ph^x_-)$ if the first alternative was $\Ph_{\me x}$ where $(+,-)$ denote the values of the spin projections along the direction chosen.  {\uf That is branch dependence.}  The relevant set of histories is thus
\begin{align}
\Ph^z_{+}U &\Ph_{\me z}, \quad \Ph^z_{-}U\Ph_{\me z}, \quad \Ph^x_{+}U\Ph_{\me x}, \nonumber \\&\quad \Ph^x_{-}U\Ph_{\me x}
\label{spin}
\end{align}
where $U\equiv \exp[-iH(t_2-t_1)/\hbar]$.
Thus branch dependent histories are needed to describe the simplest measurement when system, apparatus, and observer are all treated as subsystems of one closed system. 

{The set of histories \eqref{spin} describes all the alternatives that can happen in this limited measurement model. One could say that it is a third-person description of the possible measured histories (e.g. as in \cite{HS09}). But suppose that the result of the  observers's coin flip at $t_1$ is heads so that the spin in the $z$-direction will be measured at time $t_2$. She may be interested only in histories that describe the outcome of the projected experiment from her first-person point of view, and not the outcomes of the experiment that could have been carried out if the coin had come up tails. 
She would then use the set of histories
\be
\Ph^z_{+}U \Ph_{\me z}, \quad \Ph^z_{-}U\Ph_{\me z}, \quad  U\Ph_{\me x} 
\label{spin1}
\ee
(leaving out zero branches). This is coarser grained than \eqref{spin} but still branch dependent. }

Either way,  branch dependent sets of  histories are needed to describe realistic measurement situations when system, apparatus, and observer are all treated as subsystems of one closed system. 

\subsubsection{Planet Formation}

Quasiclassical realms have to be branch-dependent in order to have a chance of being  maximally refined with respect to decoherence and classicality (and therefore of being a feature of the universe and not our choice).

 For example, consider the formation of the Earth, starting with a protostellar cloud. A relatively coarse-grained description of the  gas might be appropriate in the
protostellar cloud, to be followed by finer and finer-grained descriptions at the locations where a star (the Sun) condensed, where a planet (the Earth) at 1AU won the battle of accretion in the circumstellar disk, etc. The higher density in the condensed region means that collision rates will be higher so the mechanisms of decoherence will operate more quickly. It also means that the same inertia is achieved in smaller volumes.
This means that the volumes of the quasiclassical realm can be smaller, and the times between alternatives can be less and still exhibit classical predictability in the face of the quantum noise produced by the mechanisms of decoherence.  The more refined set with smaller volumes and shorter times is closer to maximality. 

The locations where the Sun condensed or the Earth formed will be different on different branches. Indeed, there will be branches where they did not condense at all.  The coarse grainings described above are therefore branch dependent.

\subsection{Branch Dependence in General}
\label{branch-gen}

We now discuss branch dependence in general. It will be convenient in this Schr\"odinger picture to consider histories on a fixed time interval starting with $t_0$ and ending at $T$. (This is no restriction at all since the endpoints are arbitrary.) We can then consider histories defined at a variable number $n$ of time steps within this interval.

In a branch dependent coarse graining the quasiclassical yes/no alternatives at a given time are represented by an exhaustive set of exclusive projection operators satisfying \eqref{twoseven}.  We denote those at  time $t_k$  by\footnote{Unfortunately in previous papers we have arranged the indices differently. For instance in \cite{GH07} we wrote $P^k_{\alpha_k} (t_k;\alpha_{k-1}, \cdots, \alpha_1)$ for Heisenberg picture projections and in \cite{GH95} we wrote all the indices downstairs except $k$. These all mean the same thing. The notation used here cleanly separates the description of the set (upper indices) from the particular alternative within the set (lower index).} 
\be
\label{onetimeset}
\Ph^{k\alpha_{k-1}\cdots\alpha_1}_{\alpha_k} .
\ee
The upper indices label the set. The quantity $k$ labels both the time step and the coarse graining used at that time, as discussed in Section \ref{qcvariables}. For example we might employ  projections on a certain exhaustive set of exclusive ranges of quasiclassical variables defined by one set of averaging volumes $V(\vy)$ at one time [cf.\eqref{av-n}],  and use quasiclassical variables defined by different volumes at a different time, or different ranges $\Delta_\beta$ at different times etc. The upper indices $(\alpha_{k-1},\cdots,\alpha_1)$ indicate branch dependence --- the set of alternatives at time step $k$ depends on the previous alternatives defining a particular  history as in the examples given above. Allowing only dependence on past alternatives means that causality is built in at a basic level.  The subscript $\alpha_k$ denotes the particular alternative in the set --- a particular range of the quasiclassical variables in all the volumes. 
 
The times and the number of time steps are also branch dependent. That  must be the case if we aim at sets that are maximally refined consistent with decoherence and classicality. For instance  sets of projections can be closer together  in time when they refer to regions where decoherence is more rapid than elsewhere {\uf as in the planet formation example above}.  Thus we should write
\be
\label{times}
t_k=t_k(\alpha_{k-1}, \cdots, \alpha_1)
\ee
and a similar formula for the total number of steps $n$ in the range $t_0<t<T$.  {\cf However, in order not to expand an already complex notation, in this paper we will consider a fixed sequence of times $t_1, \cdots, t_n$ that is refined enough to accommodate all branches and use a trivial set of alternatives $(0,I)$ on particular branches where there needs to be more time separation between non-trivial alternatives.}

Histories are then represented by class operators incorporating the projections interrupted by unitary evolution.  For example with just three intermediate times we have\footnote{\cf Since the indices on the $\Ph$'s sometimes represent a set, and sometimes label an alternative, a convention has to be chosen for how they are placed on the $\Ch$'s.  This one in \eqref{classop} is  picked for later convenience.}.

\begin{align}
\label{classop}
\Ch_{\alpha_3\alpha_2\alpha_1}^3=&\Ph^{\uthree}_{\alpha_3}e^{-iH(t_3-t_2)/\hbar}\Ph^{\utwo}_{\alpha_2}e^{-iH(t_2-t_1)/\hbar} \nonumber \\  &\times\Ph^1_{\alpha_1}e^{-iH(t_1-t_0)/\hbar} .
\end{align}
 The formulae for longer sequences of times (or fewer) should be evident.

 \subsection{Adaptive Branch Dependent Coarse Grainings}
 \label{adaptive}
 Adaptive coarse grainings are branch dependent in a rule based way. For example, we may adapt the coarse graining to follow the motion of the Earth, or what happens on its surface, by choosing alternatives at a future time that describe what goes on at its future location and ignore what happens at other locations.  Appropriately adaptive coarse grainings can reduce the proliferation of branches and simplify the calculation of decoherence by focusing on histories of interest and ignoring others.  A simple example of a  general adaptive rule is, at any one time step, not to further refine branches that already have negligible probabilities. Further division of such  branches can only reduce the probabilities \cite{Harup}. A  wave packet moving in one dimension provides a very simple example.  The motion of the packet can be followed with an adaptive coarse graining that fine grains only near the center of the wave packet as it moves through successive time steps.

\subsection{Medium Decoherence of Branch Dependent Sets of Histories}
\label{med-decoh-sect}  
Assume that the universe has a pure initial state at time $t_0$ which we denote by $\Psih(t_0)$ in the Schr\"odinger picture. {\uf We will also  use $\Psih^0\equiv\Psih(t_0) $ as an alternate notation consistent with the conventions for the projections in \eqref{onetimeset}.} Consider a set of alternative histories defined by sequences of projections of the form \eqref{onetimeset} 
at a sequence of times $t_1, \cdots, t_n$ and represented by class operators \eqref{classop}. As above, to keep the notation manageable we will consider histories with just three times $t_1, t_2, t_3$.
The branch state vectors for individual histories are [cf. \eqref{twofive}]
\be
\label{branch}
\Psih^{\uthree}_{\alpha_3} \equiv \Ch_{\athree}\Psih^0 .
\ee
The sum of the branches gives back the state at time step 3, that is
\be
\label{branchsum}
\Psih^3 = \sum_{\athree} \Psih^{\uthree}_{\alpha_3},
\ee
as is easily verified from \eqref{classop} and \eqref{twoone}. 

{\cf These Schr\"odinger picture branch state vectors can be evolved to any time step  with the Hamiltonian, e.g.}
\be
\label{branch4}
\Psih^{4\alpha_2\alpha_1}_{\alpha_3} \equiv e^{-iH(t_4-t_3)/\hbar} \Ch^{3}_{\alpha_3\alpha_2\alpha_1}\Psih^0 .
\ee
Medium decoherence is the requirement that all the branches be mutually orthogonal:
\begin{align}
\label{med-decoh}
(\Psih^{\uthreep}_{\alpha'_3},& \Psih^{\uthree}_{\alpha_3}) = \delta_{\alpha'_3\alpha_3}\delta^{\alpha'_2\alpha_2}\delta^{\alpha'_1\alpha_1}p(\athree) \nonumber \\&\text{\rm (Medium Decoherence)}
\end{align}
{\uf where $p(\athree)$ are the probabilities of the histories.}

\section{Narrative Realms and Common Frameworks}
\label{narrative}

\subsection{Narrative Realms}
\label{narrative-rlms} 

Narrative realms tell a story through their probabilities  about how  features of the universe change in time. Often these stories concern the evolution in a quasiclassical realm of  unique, identifiable objects\footnote{We assume that our box is small enough that unique objects can exist in the sense that the probability for their replication elsewhere is negligible. We thus are not considering the vast universes of contemporary inflationary theory in which everything is duplicated someplace else. For what to do then see \cite{HS09}.} --- the galaxy NGC4258, the planet Mars, the Andes, eddies in your bathtub, individual human beings, and so forth. For a realm to be a narrative whose probabilities describe the history of an object the sets of projections $\{\Ph^{k\alpha_{k-1}\cdots\alpha_1}_{\alpha_k}\}$ must be chosen to that end.  At a minimum they must follow the object at different times. 

It will be useful to define narrative coarse grainings more generally than just those pertaining to objects. In general we need to capture precisely the notion that a narrative coarse graining follows similar variables at a series of times. 

The simplest example of a rule generating a narrative coarse graining is to use the same set of alternatives at all times. That is, the narrative is given by  histories that have  the same set of Schr\"odinger picture projections $\{\Ph_\alpha\}$ for all time steps. But this simple rule does not allow for branch dependence. 

A rule  more general than identity that captures the notion of similar variables at different times is to require that the Schr\"odinger picture projections commute\footnote{As discussed in Section \ref{qcvariables} for quasiclassical variables that do not commute  either we would have to find that they effectively commute as in \cite{Hal98,Hal99} or divide $k$ up into nearby separated time steps. We will not complicate an already extended notation to indicate this.}, viz. 
\be
\label{narr}
[\Ph^{k\alpha_{k-1}\cdots\alpha_1}_{\alpha_k},  \Ph^{k' \alpha'_{k'-1}\cdots\alpha'_1}_{\alpha'_k}]=0, \quad \text{\rm(Narrative Condition)}.
\ee
Evidently identical projections at different times commute, but the condition \eqref{narr} is much more general and consistent with branch dependence. We will call this the {\it narrative condition}. 

\subsection{Common Frameworks}

The narrative condition \eqref{narr} immediately leads to the notion of a common framework for narrative histories. There is a basis in $\cal H$ in which all the $\Ph$'s satisfying the narrative condition  \eqref{narr} are simultaneously diagonal. That means that there is an exhaustive set of mutually exclusive projections $\{\Ph_\gamma\}$ that is an operator basis for all the $\Ph$'s in the histories. Specifically
\be
\label{common}
\Ph^{k\alpha_{k-1}\cdots\alpha_1}_{\alpha_k} = \sum_{\gamma\in({k\alpha_{k}\cdots\alpha_1})}  \Ph_\gamma ,
\ee 
where the notation means that the $P$'s of the histories project onto orthogonal subspaces of $\Hc$ that are unions of the subspaces of the common framework.

For  quasiclassical realms one can think of the $\Ph_\gamma$ as defined by alternative values of quasiclassical variables  using a partition of space into very small volumes for averaging approximately conserved quantities   --- the same partition at all times. There is no requirement that histories of these $\Ph_\gamma$ be decoherent.  The coarse grainings for branch dependent sets of histories are defined by grouping these small volumes into appropriate larger ones in a branch dependent way that does define a decoherent set of histories. 

\section{Environments}
\label{environments}

\subsection{The Use of Environments}
\label{usesofenv}

There is a long and important history  of analyzing decoherence phenomena in terms of the interaction between a subsystem and an environment. Seminal papers in the modern quantum mechanics of closed systems include those of U. Fano \cite{Fan57}, Joos and Zeh \cite{JZ85} and the many of Zurek  and his collaborators reviewed for example in \cite{Zurekrmp}. Important earlier work, on which the present  discussion relies, includes the papers of Feynman and Vernon \cite{FV63}, Caldeira and Leggett \cite{CL83}, and our own \cite{GH93a}.  In these treatments one set of fundamental coordinates (say $x$'s)  define the subsystem, the rest (say $Q$'s) define the environment. For instance, the subsystem might be a single dust particle interacting with photons of the cosmic background radiation radiation constituting an environment. Corresponding to this division of coordinates there is a tensor product factoring of the Hilbert space $\ch$ 
\begin{equation}
\ch ={\ch^s} \otimes {\ch^e} 
\label{tensor}
\end{equation}
with $\ch^s$ spanned by center of mass position of the dust grain  and $\ch^e$ by the field variables of the photons and the internal relative coordinates of the atoms in the grain. The Hilbert space $\ch^s$ is for the `system' and $\cH^e$ defines its `environment'. 

The ubiquity of models assuming a system-environment split has given some the impression that such a split must be postulated to formulate the quantum mechanics of closed systems. This is not correct.

When quantum mechanics is formulated generally there is no fundamental system-environment split. The general notion is rather coarse graining. As mentioned above, the most general objective of quantum theory is the prediction of the probabilities of sets of alternative coarse-grained histories of a closed system. Coarse-graining is inevitable because there are no non-trivial fine-grained sets of histories that decohere. It is the coarse graining that specifies what is followed and what is ignored. No additional separate postulate of a system-environment split is needed to extract predictions from the theory. Whatever notion of system and environment may be available follows from the coarse graining defining a particular set of alternative histories and will differ from one set of alternative histories to another and indeed within the set as well. Environments are not postulated; they are constructed from sets of sets of histories. We now describe how that works. 

\subsection{Constructing Environments}
\label{constructing}

First, let's consider branch independent realms. We will return to the more general branch dependent case immediately afterwards.
 
For a given branch-independent realm a system-environment split is defined  at one time when the Hilbert space $\cH$  is a tensor product as in \eqref{tensor}
 and all  the projections defining the histories are  of the form
\begin{equation}
\Ph_\alpha = \Ph_\alpha^s \otimes I^e 
\label{tensorp}
\end{equation}
at that time for that realm.
The Hilbert space $\ch^s$ is for the `system' and $\cH^e$ defines its `environment'.

The important mathematical result is the converse. 
Given a set of commuting projection operators $\{P_\alpha\}$, $\alpha=1,2,\cdots$,  all defining infinite-dimensional subspaces of $\ch$, it is possible to factor the  $\ch$ as in \eqref{tensor} so that all the projections act on one factor as in \eqref{tensorp}.  The argument is a simple one\footnote{See Appendix A of \cite{BH99}. There is an obstruction to factorization in the finite-dimensional case arising from the relation between dimensions following from \eqref{tensorp}, specifically: $\dm(P_\alpha) = \dm(P^s_\alpha)\dm(I^e)$. }. Since all the projections commute they can be written in a common basis $\{|i\rangle\}$in the form
\begin{equation}
\Ph_\alpha = \sum_{i\in\alpha} |i\rangle\langle i | .
\label{sumi}
\end{equation}
Then  it's is just a matter of relabeling the basis $|i\rangle\equiv |\alpha, A\rangle$ to define a tensor product \eqref{tensor} on which the projections act as in \eqref{tensorp}. 

In the simplest case its is possible to arrange for the projections on the system space to be one dimensional
\be
\Ph^s_\alpha \equiv |\alpha\rangle^{s} \  {^s}\langle \alpha|  . 
\label{1-d}
\ee
The system Hilbert space  $\cH^s$ is spanned by all the $|\alpha\rangle^s$'s and the environment Hilbert space $\cH^e$ is spanned by all the $|A\rangle^e$'s. The environment Hilbert space is then as large as possible allowing the most Hilbert space in which the phases between histories can get lost. That is the best possibility  for decoherence. 

But it can be  convenient to allocate a little more of the Hilbert space to the system by assigning some of the $|A\rangle^e$'s to $\cH^s$. Then the system Hilbert space $\cH^s$ is spanned by vectors $|\alpha, r\rangle^s$ and 
\be
\Ph^s_\alpha = \Sigma_r |\alpha, r \rangle^s \ {^s}\langle r, \alpha|  . 
\label{claw}
\ee
Thus, in a set of histories defined by sequences of sets of branch {\it in}dependent  commuting projections $\{\Ph^k_{\alpha_k}\}$, there would be a system-environment  split defined at each moment of time although generally a different split from one moment to the next. For different realms defined by different sets of projections there would be different splits.

\subsection{System-Environment Splits for Branch Dependent Coarse Grainings}
\label{SEbranch}

 In the more realistic branch dependent case we do not generally have one set of commuting projections at each time. The branch dependent projections \eqref{onetimeset} need not commute for different values of $\alpha_{n-1}, \cdots \alpha_1$. The simple measurement situation described in Section \ref{measurement}  is an example. For a general set of branch dependent histories there will be no notion of environment available even at one time. 
 
 A system-environment split of the Hilbert space can be constructed at one time when there is something in common that is followed by all the projections $\{\Ph^{k\alpha_{k-1}\cdots\alpha_1}_{\alpha_k}\}$. That can constitute the `system' and the rest is the `environment'.  Mathematically this idea is implemented when all the projections at a given  time commute, viz. 
\be
\label{commeqtime}
[\Ph^{k\alpha_{k-1}\cdots\alpha_1}_{\alpha_k},  \Ph^{k \alpha'_{k-1}\cdots\alpha'_1}_{\alpha'_k}]=0,
\ee
This  is of the same form as the narrative condition \eqref{narr} but enforced only when the times are the same.  As in that discussion, there is now an operator basis for all the $\Ph$'s and we can write for each time step $k$
\begin{equation}
\Ph^{k\alpha_{k-1}\cdots\alpha_1}_{\alpha_k}= \sum_{\gamma \in  (\alpha_{k}\cdots\alpha_1)}\Ph^k_{\gamma}\quad\quad  ({\rm all} \ \ \alpha_{k-1}, \cdots, \alpha_1)
\label{common}
\end{equation}
where the sum is over all $\Ph^k_\gamma$ contained in the projection $\Pb$. The common framework $\Ph^k_\gamma$ can then be used to factor the Hilbert space as in Section \ref{constructing}  and define a system-environment split at each time.  If the common frameworks  $\{\Ph^k_\gamma\}$ are the same for all times (all $k$) then an environment can be defined that is fixed for all time. Sets of histories constructed from the $\Ph^k_\gamma$  themselves are not necessarily decoherent. Rather they provide a  common framework for the branch dependent sets that do decohere. 

When a branch dependent set of  histories is a narrative so that \eqref{narr} is satisfied, then there is a common framework for all times
\eqref{common} and one system-environment split for all times.  That will be the case  for quasiclassical realms since we define them to be narratives.  

\subsection{Constructing Common Frameworks}
\label{constrcommon}

For a given time step $k$ the $\Ph^k_\gamma$ can be constructed from the projections $\Pb$ when these all commute with one another as in \eqref{commeqtime}. To see this
 consider for simplicity $k=2$ and define the operator products
\be 
{\mathring P}^{2\alpha_1\alpha'_1}_{\alpha_2\alpha'_2} \equiv \Ph^{2\alpha_1}_{\alpha_2}\Ph^{2\alpha'_1}_{\alpha'_2}.
\label{intersect}
\ee
Since the the $\Ph^{2\alpha_1}_{\alpha_2}$ commute for with each other for different indices by assumption \eqref{commeqtime}, the ${\mathring P}$'s are themselves projectors and the set of them an exhaustive set of exclusive projections.  In fact, ${\mathring P}^{2\alpha_1\alpha'_1}_{\alpha_2\alpha'_2}$ projects on the intersection of the subspaces defined by its constituent projections. The $\Ph$'s can be recovered from the $\mathring P$'s by, e.g.  
\be
\label{revcov}
\Ph^{2\alpha_1}_{\alpha_2}= \sum_{\alpha'_2\alpha'_1} {\mathring P}^{2\alpha_1\alpha'_1}_{\alpha_2\alpha'_2}
\ee
Thus, the $\Ph^2_\gamma$'s in \eqref{common}  are the  ${\mathring P}^{2\alpha_1\alpha'_1}_{\alpha_2\alpha'_2}$'s. The index $\gamma$ ranges over the intersections of the $\Ph^{2\alpha_1}_{\alpha_2}$'s. 

This explicit construction becomes increasingly complex for later times because all possible products of projections defining the histories enter. The idea is the same; the equations become lengthy.

We can therefore have a basis in $\cH$ of the form $|\gamma, B\rangle$ where $\gamma$ labels the intersections and $B$ labels the vectors in the intersections. We can then invoke the arguments in Section \ref{constructing} to make a system-environment split  where the system Hilbert space $\cH^s$ is spanned by vectors $|\gamma\rangle$ and the environment Hilbert space by the $|B\rangle$'s.  

{\uf We now use these ideas to define strong decoherence.}


\section{Strong Decoherence Simplified}
\label{strong}
{\uf As mentioned in the Introduction a wide class of realistic mechanisms of decoherence are characterized by the creation of orthogonal records leading to a notion of decoherence which we have called strong decoherence \cite{GH95,Fin93}. }
Section \ref{environments}'s construction of environments for each time step from a common framework permits a simplified but general  discussion of strong decoherence. We present that in this section.  For simplicity we give the exposition for histories with just three time steps, but the generalization of the formulae to more (or fewer) steps should be evident. 

The assumption that all the $\Ph$'s at a given time commute  \eqref{commeqtime} allows a system-environment split at each time as discussed in Section \ref{environments}.  At time step 3 we would have 
\be
\label{tensor3}
\Hc = \Hc^{3s} \otimes \Hc^{3e} ,
\ee
where the $\Ph$'s operate only on the system Hilbert space,  $\Hc^{3s}$ at time step 3. Explicitly this means 
\be
\label{v-def}
\Ph_{\alpha_3}^{\uthree} = \sum_{r_3} v^{\uthree}_{\alpha_3 r_3}v^{\dagger\uthree}_{\alpha_3 r_3} \otimes I^{3e}
\ee
where  the $v$'s are a set of orthogonal  basis vectors in $\Hc^{3s}$ which can be arranged to satisfy
\be
\label{v-ortho}
(v^{\uthree}_{\alpha_3 r_3},v^{\uthree}_{\alpha'_3 r'_3})^{3s}=\delta_{\alpha'_3\alpha_3}\delta_{r'_3 r_3},
\ee
Since the $\Ph$'s are an exhaustive set of projections, the set of $v$'s for all projections  in the set (all $\alpha_3)$ will be a basis for $\cH^{3s}$ and we can expand $\Psih^{\uthree}_{\alpha_3}$ in terms of them, viz. 
\be
\label{psi-v-z}
\Psih^{\uthree}_{\alpha_3} = \sum_{r_3} v^{\uthree}_{\alpha_3 r_3} \otimes z^{\uthree}_{\alpha_3r_3} .
\ee
The coeffecients $z^{\uthree}_{\alpha_3r_3}$ are vectors in the environment Hilbert space at time step  $3$, ${\cal H}^{3e}$. 

The condition for medium decoherence \eqref{med-decoh} then becomes
\begin{align}
\label{med-decoh-v-z}(\Psih^{\uthree}_{\alpha_3},& \Psih^{\uthreep}_{\alpha'_3})  = \sum_{r_3r'_3}(v^{\uthree}_{\alpha_3 r_3},v^{\uthreep}_{\alpha'_3 r'_3})^s \nonumber \\
&\times(z^{\uthree}_{\alpha_3 r_3},z^{\uthreep}_{\alpha'_3 r'_3})^e  \propto \delta_{\alpha'_3\alpha_3}\delta^{\alpha'_2\alpha_2}\delta^{\alpha'_1\alpha_1}.
\end{align}
 The idea of strong decoherence is that we require orthogonality of the $z$'s in the past alternativesm,
viz
\be 
\label{strong-z}
(z^{\uthree}_{\alpha_3 r_3},z^{\uthreep}_{\alpha'_3 r'_3})^{3e}  \propto \delta^{\alpha_2\alpha'_2}\delta^{\alpha_1\alpha'_1},   \quad \text{\rm (Strong Decoherence)}
\ee
for all $\alpha_3$, $\alpha'_3$, $r_3$, $r'_3$. 
Note that we don't require orthogonality in the the index $\alpha_3$. There is no need for it. 
Orthogonality in $\alpha_3$ is automatic from \eqref{v-ortho} once the \eqref{strong-z} is satisfied.  Further, as we will see in the next section the $z$'s are connected with records of the histories, and physically it takes some time for records to form. Non-orthogonality in $\alpha_3$ is consistent with that. 

It is easy to see that strong decoherence is a stronger condition than medium decoherence. Strong decoherence ensures that 
\be
\label{stronger}
(\Psih^{\uthree}_{\alpha_3}, {\chO}\Psih^{\uthreep}_{\alpha'_3}) \propto \delta^{\alpha_2\alpha'_2}\delta^{\alpha'_1\alpha'_1}
\ee
for any operator $\chO$ of the form 
\be
\label{systobs}
{\chO}^{3s}={\chO}^{3s}\otimes I^{3e}
\ee
not just for ${\chO}^{3s}=I^{3s}\otimes I^{3e} = I$ which is all that medium decoherence ensures. 

As defined here strong decoherence requires only that the projections defining the branches commute at each time \eqref{commeqtime}. There is a system-environment split for each time, but the split generally  changes from one time to the next.
When the set of histories is a narrative all the projections at different times commute \eqref{narr}. There is then a common framework for all projections at all times \eqref{common}, and correspondingly a notion of a system-environment split for all times. The notion of system --- that which is followed in common by all the projections --- will necessarily be more restricted than it is at one time unless the projections at different times are connected by a very simple narrative rule (e.g identity).  That is because there will be many more intersections to consider in the construction of the common framework as in Section \ref{constrcommon}. That does not make strong decoherence easier, but it would enable system and environment to be followed separately over time as will be described in the next section. 

We now turn to describing some consequences of this strong decoherence condition.

\section{Consequences  of Strong Decoherence}
\label{features} 

\subsection{Records}
\label{records}
A record of a history is an alternative at one time that {\uf has a high probability of correlation  with} alternatives in the history at an earlier time. A set of alternative histories is said to be {\it recorded} if there is a set of alternatives at one time one of which is correlated with each past history in the set.  To see how this idea is implemented in the present framework we continue with just two  time steps represented by class operators 
\be
\label{classop2}
\Ch_{\alpha_2\alpha_1}^2=\Ph^{\utwo}_{\alpha_2}e^{-iH(t_2-t_1)/\hbar}\Ph^1_{\alpha_1}e^{-iH(t_1-t_0)/\hbar} .
\ee
The illustration with this simple case should be sufficient to see how to generate more general formulae with more steps. 

This set  of histories represented by \eqref{classop2} is recorded at time step $t_3$  if there is a set of commuting,  orthogonal projections $\{\R3\}$ satisfying  [cf. \eqref{twoone}]
\begin{subequations}
\label{records-def}
\be 
\Rh^3_{\atwo} \Rh^3_{\atwop} = \delta_{\alpha_1\alpha'_1}\delta_{\alpha_2\alpha'_2}\Rh^3_{\atwo}
\label{record-ortho}
\ee
such that 
\be
\label{record4}
\R3 \Psih^3  = \Psih^{3\alpha_1}_{\alpha_2} \equiv e^{-iH(t_3-t_2)/\hbar} \Ch^{2}_{\alpha_2\alpha_1}\Psih^0 .
\ee
\end{subequations}
Taking a time step after the last one in the histories captures the idea that it might take time for a record to form. 

As a consequence of strong decoherence there are always records of past history in the environment satisfying \eqref{records-def}. Examples can be exhibited explicitly.  In the environment Hilbert space $\cH^{3e}$ at time step 3 define the following set of projections:
\begin{subequations}
\label{zrecords}
\be
\Rh^{3e}_{\atwo} = \text{\sf Proj}(z^{3\atwo}_{\alpha_3 r_3})
\label{record-def}
\ee
and
\be
\Rh^{3}_{\atwo} \equiv {I^s} \otimes {\Rh}^{3e}_{\atwo} .
\label{rec-full}
\ee
\end{subequations} 
Here, {\sf Proj} means projections on the  subspace of $\cH^{3e}$ spanned by $z^{3\atwo}_{\alpha_3 r_3}$ as $\alpha_3$ and $r_3$  vary. 
It is then a straightforward calculation using the strong decoherence condition \eqref{strong-z} to verify that  with  these definitions the record conditions \eqref{records-def} are satisfied. Since the $z$'s are generally not a basis in the environment Hilbert space, there will generally be other choices of record operators satisfying \eqref{records-def} containing those in \eqref{zrecords}. 

Note that were the analogous construction of $R$'s made at $t_2$, the last time of the history, it would not have worked. Mathematically that is because the strong decoherence condition on the $z$'s at that time would not have ensured orthogonality in $\alpha_2$. That is consistent with the physical idea that records of alternatives are not available instantaneously but generally take some time to form. 

Many mechanisms of decoherence that have been studied in simple models involve a coupling of a followed system to an environment of different degrees of freedom. The environmental degrees of freedom carry away the phases between alternative histories of the followed degrees of freedom and produce decoherence. After the interaction the environmental degrees of freedom contain records of the configuration of the followed system at the time of interaction\footnote{See e. g. \cite{Hal99a} for  records in the oscillator models. For an emphasis on the redundancy of records see e.g. \cite{darwinism}. }.  Both environments and records in environments are consequences of strong decoherence as we have seen in this section and in Section \ref{environments}. Restricting quasiclassical realms  to strongly decoherent histories of quasiclassical variables thus captures, in a general way, key features of realistic mechanisms of decoherence.

\subsection{Permanence of the Past}
\label{past}

We experience the present, remember the past, and try to predict the future. We have the impression that the future is uncertain, waiting to happen. By contrast, the past is over, done with, and {\uf permanent} even if our knowledge of what happened is uncertain. 
But these subjective ways of organizing temporal information and these impressions are not built into the fundamental laws of the quantum universe. Plausibly they  rather arise from our particular construction as physical systems within the universe\footnote{That is as an IGUSes --- an information gathering and utilizing system \cite{Gel94}.} \cite{Har05b}. At every moment of time in our history of  there is a present separating a past from a future.

Consider the present, past and future at a particular moment in our history.  In decoherent histories quantum theory there is no essential  difference between using present data to predict the future and using it to  retrodict the past \cite{GH90a,Har98b}. Both prediction and retrodiction involve the probabilities of histories conditioned on present data --- one of histories that extend toward the big bang (the past) and the other away from it (the future)\footnote{Although the formulae for these probabilities differ in form \cite{GH90a}.}. One gives probabilities of what did happen, the other of what will happen.

There may be many realms extending the present towards the future and many towards the past. Neither past or future is therefore unique \cite{Har98b}. However, usually we are concerned with the past or future of a quasiclassical realm. We will assume that here.

Extending a quasiclassical realm into the future risks losing the ability to retrodict the past. That is because any extension into the future is a fine-graining of the present set of alternative histories. A coarse-graining of a decoherent set of alternative histories is decoherent. 
A fine-graining may not be. Extending a realm to the future risks losing the decoherence of the past. The past is  therefore not necessarily permanent (e.g. \cite{Har98b}).

Strong decoherence ensures the permanence of the past. That is because the condition \eqref{strong-z} requires the decoherence of past alternatives. A more physical way of saying this is that, as discussed above,  strong decoherence ensures the existence of present records for the past that ensure its decoherence and permanence. 

\subsection{Density Matrices}
\label{denmatrix}

When there is a system-environment split of the Hilbert space at any one time $k$  as in \eqref{tensor3} it is possible to define a system density matrix $\rho^{ks}$ by tracing over the environment. Specifically,
\be
\rho^{ks} = Sp(\Psih^k{\Psih^{k^\dagger}})
\label{rhosyst}
\ee
where $Sp$ means the trace over the environment Hilbert space $\cH^{ke}$. The expected value of any system observable of the form \eqref{systobs} at time step $k$ can be calculated just from $\rho^{ks}$, viz
\be
\label{expect-syst}
\langle {\cO} \rangle^k \equiv Tr({\cO} \Psih^k\Psih^{k\dagger}) = tr(\cO^s \rho^{ks}) .
\ee
where $Tr$ is the trace over all of Hilbert space $\cH$ and $tr$ is the trace over the system part ${\cH^{ks}}$.  

In this section we show that strong decoherence implies this result on a branch by branch basis. Specifically we show the following:  Define at  time step $k$, for each branch $\branch$, its  branch density matrix in ${\cH^{ks}}$.  

\be
\label{branch-dm}
\rho^{\branchk s}  \equiv Sp(\Psih^{k\branch}\Psih^{\branchk \dagger}) . 
\ee
Then if $\cO$ is a system observable of the form \eqref{systobs} for the system-environment split at time step $k$,  strong decoherence implies
\begin{align}
\label{expected-k}
\langle \cO \rangle^k &\equiv Tr(\cO\Psih^k\Psih^{k\dagger})  = tr({\cO^s}  {\rho^{ks}}) \nonumber \\
&= \sum_\branch tr({\cO^s} {\rho^{\branchk s}}) . 
\end{align}

We illustrate the demonstration with just three time steps as in Sections \ref{med-decoh-sect} and \ref{strong}. The generalization to more steps should be straightforward. We begin by using \eqref{psi-v-z} to write the expected value of a system observable \eqref{systobs} at time step 3 in terms of the $v$'s and $z$'s,
\begin{align}
\label{expect-vz}
\langle\cO\rangle^3 &\equiv (\Psih^{3},\cO^s\Psih^3) \nonumber \\
&=\sum_{\alpha_3\alpha_2\alpha_1}\sum_{\alpha'_3\alpha'_2\alpha'_1}(v^{\uthree}_{\alpha_3 r_3},{^s\chO^s}v^{\uthreep}_{\alpha'_3 r'_3})^s
(z^{\uthree}_{\alpha_3 r_3},z^{\uthreep}_{\alpha'_3 r'_3})^e \nonumber \\
&=\sum_{\alpha_2\alpha_1}\sum_{\alpha_3\alpha'_3}(v^{\uthree}_{\alpha_3 r_3},{^s\chO^s}v^{\uthree}_{\alpha'_3 r'_3})^s
(z^{\uthree}_{\alpha_3 r_3},z^{\uthree}_{\alpha'_3 r'_3})^e
\end{align}
The last  equality is a consequence of the strong decoherence condition \eqref{strong-z}. The result gives the expected value of a system observable  $\langle\cO\rangle^3$ as a single sum over branches $\alpha_2,\alpha_1$. To put it differently, strong decoherence means that there is no interference between branches. Expanding \eqref{branch-dm} in a similar way in terms of the $v$'s and $z$'s shows that the matrix elements of $\rho^{\branchk s}$ in the basis of $\{ v^{\uthree}_{\alpha_3 r_3}\}$'s  in the system Hilbert space $\cH^{3s}$ are $(z^{\uthree}_{\alpha_3 r_3},z^{\uthree}_{\alpha'_3 r'_3})^e$.  Equation \eqref{expect-vz} is then \eqref{expected-k} in this particular basis. 

We note that diagonality of the density matrices $\rho^{\branchk s}$ is {\it not} a consequence of strong or medium deocherence. 
Diagonality would mean that 
\be
\label{toostrong}
(z^{\uthree}_{\alpha_3 r_3},z^{\uthreep}_{\alpha'_3 r'_3})^{3e}  \propto \delta_{\alpha_3\alpha'_3}\delta^{\alpha_2\alpha'_2}\delta^{\alpha_1\alpha'_1} \quad  \text{\rm (Too Strong)}
\ee
which is a stronger condition than strong decoherence\footnote{The condition is so strong that it would imply  that the expansion \eqref{expect-vz}  is a Schmidt decomposition, which would fix the variables in the projections and risk conflict with our assumption that they are quasiclassical.}.  This condition is not necessary for medium decoherence. The medium decoherence condition \eqref{med-decoh} is satisfied in $\alpha_3$ automatically as a consequence of the orthogonality of the 
$v$'s as \eqref{med-decoh-v-z} shows. Further, strong decoherence is equivalent to the creation of records in the environment as  we showed in Section 
\ref{records}. One would expect that physical records would not appear simultaneously with the alternative but take some time to form.

Models of decoherence may lead to a density matrix becoming diagonal after time\footnote{For example as in some oscillator models. See, e.g. \cite{Har91a}, Section II.6.4} but that plays no role in the fundamental formulation of decoherent histories quantum theory. 

\section{Conclusion}
\label{conclusion}
If the universe is indeed a quantum mechanical system then at a fundamental level the  predictions of theory are the probabilities of the individual members of  sets of alternative coarse-grained histories ---  realms. Of particular interest are the quasiclassical realms that  are a feature of our universe, extending over the whole {\uf of its visible part}  from just after the big bang to the far future. These describe almost everything we observe from every day scales to those of cosmology. Characterizing the universe's quasiclassical realms is therefore an important problem in quantum mechanics.

This paper has continued a program of characterizing the quasiclassical realms in decoherent histories quantum theory when quantum gravity is neglected and classical spacetime is assumed. This paper discussed the ideas of adaptive branch dependent coarse grainings, narrative sets of histories, and strong decoherence. Putting together all the elements of this paper and our previous ones, we can characterize a quasiclassical realm as a strongly decoherent set of alternative histories, defined by an adaptive branch dependent coarse graining built on a narrative framework of quasiclassical variables,  exhibiting with high probability patterns of correlation in time described by closed sets of deterministic equations, and maximally refined consistent with all these properties. 

\appendix
\renewcommand{\theequation}{\Alph{section}.\arabic{equation}}

\section{Quasiclassical Coarse Grainings} 
\label{appA}

In this appendix we describe explicitly the quasiclassical coarse grainings that are implicitly referred to in the previous sections. 

Begin by partitioning the interior of the box containing our model universe into regions labeled by a discrete triple of indices $\vy$ with spatial volumes $V(\vy)$. Denote by $\ep(\vec x)$, $\pv(\vec x)$ and $\num(\vec x)$ the Schr\"odinger picture operators for energy density, momentum density, and  number density respectively. (In this appendix we suppress the hats on these quantities so that the notation doesn't become too  messy.) The averages of the energy density  over the volumes are then then defined by
\begin{equation}
\bar\epsilon (\vec y)  \equiv  \frac{1}{V(\vec y)} \int_{V(\vec y)} d^3x\, \epsilon (\vec x) .
\label{av-n}
\end{equation} 
with similar expressions for the other quantities. 

A coarse-grained description of the value $\bar\epsilon(\vec y)$  is provided by a partition of the real line into an exhaustive set of exclusive ranges $\Delta_\beta$, $\beta =1,2,\cdots$. To ask for the coarse-grained value of $\bar\epsilon(\vy)$ is to ask whether it lies in the range $\Delta^{\vy}_\beta$ --- yes or no --- for all $\beta$. These alternatives correspond to projection operators ${\hat P}^{\bar\epsilon(\vec y)}_{\beta}$. Histories of the values of $\bar\epsilon(\vy)$ would be represented by sequences of such projections  --- one for each $\vy$ at each time step as in \eqref{classop-simp}. 

Including the other two variables would involve products of projectors like ${\hat P}^{\bar\epsilon(\vec y)}_{\beta}$ for the other quantities. Since since they generally do not commute with each other the time steps may have to be slightly separated, or their effective commutation established as in \cite{Hal98,Hal99}.

\renewcommand{\theequation}{\Alph{section}.\arabic{equation}}

\acknowledgments
We thank the Aspen Center for Physics for hospitality over several summers while this work was in progress. JBH thanks the Santa Fe Institute for supporting many visits there. The work of JBH was supported in part by the National Science Foundation under grants PHY05-55669, PHY08-55415, and PHY12-05500.
 The work of MG-M was supported by the C.O.U.Q. Foundation, by Insight Venture Management, and by the KITP in Santa Barbara.  The generous help provided by these organizations is gratefully acknowledged.


\begin{thebibliography}{99}


\bibitem{GH90a} M.~Gell-Mann and J.B.~Hartle, {\it Quantum Mechanics in the
Light of Quantum Cosmology}, in {\sl Complexity, Entropy, and the Physics
of Information, SFI Studies in the Sciences of Complexity}, Vol.~VIII,
ed.~by W.~Zurek,  Addison Wesley, Reading, MA (1990).

\bibitem{GH90b} M.~Gell-Mann and J.B.~Hartle, {\it Alternative Decohering
Histories in Quantum Mechanics}, in the {\sl Proceedings of
the 25th International Conference on High Energy Physics}, Singapore,
August, 2-8, 1990, ed.~by K.K.~Phua and Y.~Yamaguchi (South East Asia
Theoretical Physics Association and Physical Society of Japan) distributed
by World Scientific, Singapore (1990).

\bibitem{Har91a} J.B.~Hartle, {\it The Quantum Mechanics of Cosmology}, in {\sl
Quantum Cosmology and Baby Universes:  Proceedings of the 1989 Jerusalem Winter
School for Theoretical Physics}, ed.~by ~S.~Coleman, J.B.~Hartle, T.~Piran,
and S.~Weinberg, World Scientific, Singapore (1991), pp. 65-157.

\bibitem{GH93a} M.~Gell-Mann and J.B.~Hartle, {\it Classical Equations for
Quantum Systems}, {\sl Phys.~Rev.~D}, {\bf 47}, 3345 (1993);
arXiv:gr-qc/9210010.

\bibitem{GH93b}  M.~Gell-Mann and J.B.~Hartle, {\it Time Symmetry and
Asymmetry in Quantum Mechanics and Quantum Cosmology}, in {\sl The Physical
Origins of Time Asymmetry}, ed.~by J.~Halliwell, J.~P\'erez-Mercader, and W.~Zurek,
Cambridge University Press, Cambridge (1994); arXiv:gr-qc/9309012.

\bibitem{Gel94} M.~Gell-Mann, {\sl The Quark and the Jaguar}, W.H..~Freeman
New York (1994).

\bibitem{GH94} M.~Gell-Mann and J.B.~Hartle, {\it Equivalent Sets of
Histories and Multiple Quasiclassical Domains}; arXiv:gr-qc/9404013.

\bibitem{Har94b} J.B.~Hartle, {\it Quasiclassical Realms in A Quantum Universe},
 in {\sl Proceedings of the Cornelius  Lanczos International
 Centenary Conference}, ed.~by J.D.~Brown, M.T.~Chu, D.C.~Ellison, R.J.~Plemmons,
SIAM, Philadelphia, (1994); arXiv:gr-qc/9404017.

\bibitem{Har09} J.B.~Hartle, {\sl The Quasiclassical Realms of this Quantum Universe,} {\sl Foundations of Physics}, {\bf 41}, 982 (2011);  arXiv:0806.3776. A slightly shorter version is published in  {\sl Many Quantum Worlds} edited by A. Kent and S. Saunders (Oxford University Press, Oxford, 2009). 

\bibitem{GH95}  M.~Gell-Mann and J.B.~Hartle, {\it Strong Decoherence},
in the {\sl Proceedings of the 4th Drexel Symposium on
Quantum Non-Integrability --- The Quantum-Classical Correspondence},
Drexel University, September 8-11, 1994, ed.~by. D.-H.~Feng and B.-L.~Hu,
International Press, Boston/Hong-Kong (1995); arXiv:gr-qc/9509054.

\bibitem{GH07} M. Gell-Mann and J.B~Hartle, {\it Quasiclassical Coarse Graining and Thermodynamic Entropy}, {\sl Phys. Rev. A}, {\bf 76}, 022104 (2007), arXiv:quant-ph/0609190.

\bibitem{Gri02} R.B.~Griffiths, {\sl Consistent Quantum Theory}, Cambridge
University Press, Cambridge, UK (2002).

\bibitem{Omn94} R.~Omn\`es, {\sl Interpretation of Quantum Mechanics},
Princeton University Press, Princeton (1994).
 
\bibitem{Har95c} J.B.~Hartle, {\it Spacetime Quantum Mechanics and the
Quantum Mechanics of Spacetime} in {\sl Gravitation and Quantizations},
 {\sl Proceedings of the 1992 Les Houches Summer School},
ed.~by B.~Julia and J.~Zinn-Justin, Les Houches Summer
School Proceedings, Vol.~LVII, (North Holland, Amsterdam, 1995);
arXiv:gr-qc/9304006. A pr\'ecis of these lectures is given in {\it Quantum Mechanics at the Planck
Scale}, talk given at the {\sl Workshop on Physics at the Planck Scale}, Puri,
India, December 1994; arXiv:gr-qc/9508023.

\bibitem{Har06} J.B.~Hartle,  {\it Generalizing Quantum Mechanics for Quantum Spacetime} in {\sl The Quantum Structure of Space and Time, Proceedings of the 23rd Solvay Conference on Physics}, ed. by D. Gross, M. Henneaux, and A. Sevrin (World Scientific, Singapore, 2007), arXiv:gr-qc/0602013.

\bibitem{STalt}  For small subset of  the possible examples of this see e.g., J.B.~Hartle, {\it Spacetime Coarse Grainings in
Non-Relativistic QuantumMechanics}, {\sl Phys. Rev., D}  {\bf 44}, 3173 (1991); D.~ Marolf, {\it Almost Ideal Clocks in Quantum Cosmology: A Brief Derivation of Time}, {\sl Class. Quant. Grav.} {\bf 12}, 2469 (1995); arXiv:gr-qc/9412016;  J.~Halliwell, {\it  Quantum Cosmological Models: A Decoherent Histories Analysis Using a Complex Potential}, {\sl Phys. Rev. D}, {\bf 80}, 124032 (2009), arXiv:0909.2597. 

\bibitem{Dio04}  L.~Di\'osi, {\it Anomalies of Weakened Decoherence Criteria for Quantum Histories}, {\sl Phys.~Rev.~Lett.}, {\bf 92}, 170401 (2004). 

\bibitem{Hal98} J.~Halliwell, {\it Decoherent Histories and Hydrodynamic Equations},
{\sl Phys.~Rev.~D}, {\bf 58}, 1050151--12 (1998); arXiv:quant-ph/9805062.

\bibitem{Hal99}J.~Halliwell, {\it  Decoherent Histories and the Emergent Classicality of Local Densities}, {\sl  Phys. Rev. Lett.}  {\bf 83},  2481 (1999); arXiv:quant-ph/9905094.

\bibitem{HS09} M.~Srednicki and J.B.~Hartle, {\it Science in a Very Large Universe}, {\sl Phys. Rev. D}, {\bf 81} 123524 (2010),  arXiv:0906.0042, {\it The Xerographic Distribution: Scientific Reasoning in a Large Universe}, arXiv:1004.3816. 

\bibitem{Harup} J.B.~Hartle, unpublished. 

\bibitem{Fan57}  U.~ Fano, {\it Description of States in Quantum Mechanics by Density Matrix and Operator Techniques}, {\sl Rev. Mod. Phys}, {\bf 29}, 74 (1957). 

\bibitem{JZ85}  E.~Joos and H.D.~Zeh, {\it The Emergence of Classical Properties Through
Interaction with the Environment}, {\sl Zeit. Phys.~B}, {\bf 59}, 223 (1985).

\bibitem{Zurekrmp}  Wojtek W.~Zurek, {\it Decoherence, Einselection, and the Quantum Origin of the Classical}, {\sl Rev. Mod. Phys.}, {\bf 75}, 715 (2003), arXiv:quant-ph/0105127. 

\bibitem{FV63}  R.P.~Feynman and J.R.~Vernon, {\it The Theory of a General Quantum System
Interacting with a Linear Dissipative System}, {\sl Ann.~Phys.~(N.Y.)}
{\bf 24}, 118--173 (1963).

\bibitem{CL83} A.~Caldeira  and A.~Leggett, {\it Path Integral Approach to Quantum
Brownian Motion}, {\sl Physica A} {\bf 121}, 587 (1983).

\bibitem{BH99} T.~Brun and J.B.~Hartle, {\it Classical Dynamics of the
Quantum Harmonic Chain}, {\sl Phys.~Rev.~D}, {\bf 60}, 123503 (1999);
arXiv:quant-ph/9905079. 

\bibitem{Fin93} J.~Finkelstein, {\it Definition of Decoherence}, {\sl Phys. Rev. D}, {\bf 47}, 5430 (1993). 

\bibitem{Hal99a} J.~Halliwell, {\it Somewhere in the Universe: Where is the
Information Stored When Histories Decohere?}, {\sl Phys.~Rev.~D} {\bf 60},
105031 (1999); arXiv:quant-ph/9902008.

\bibitem{darwinism}  See e.g,  W~Zurek, {\it Quantum Darwinism}, {\sl Nature Physics}, {\bf 5}, 181 (2009), 
C.J.~Reidel and W.~Zurek, {\sl The Objective Past of a Quantum Universe, Pt 1: Redundant Records of Consistent Histrories}, arXiv:1312.0331.

\bibitem{Har05b} J.B~Hartle, {\it  The Physics of Now}, {\sl Am. J. Phys.}, {\bf 73}, 101-109 (2005), arXiv:gr-qc/0403001.

\bibitem{Har98b} J.B.~Hartle, {\it Quantum Pasts and the Utility of History}, {\sl Physica
Scripta}, {\bf T76}, 67--77 (1998); arXiv:gr-qc/9712001.

\bibitem{Dio95} L.~Di\'osi, {\it On the Maximum Number of Decoherent Histories}, {\sl Phys. Lett.}, {\bf 203}, 267-268 (1995); arXiv:gr-qc/9409028.

\end{thebibliography}
\end{document}